\documentclass[12pt]{article}

\usepackage{scicite_modified}
\usepackage[labelfont=bf]{caption}
\usepackage[normalem]{ulem}
\usepackage{xcolor}
\usepackage{times}
\usepackage[]{graphicx}
\usepackage{setspace}
\usepackage{amssymb}
\usepackage{amsmath}
\usepackage{bm}

\usepackage[outer=1.75cm,inner=1.75cm,top=2.8cm,bottom=2.8cm]{geometry}

\newcounter{lastnote}

\title{Physical Limits on Bacterial Navigation in \\ Dynamic Environments} 

\author
{Andrew M. Hein$^{1,*}$,  Douglas R. Brumley$^{2,3}$, Francesco Carrara$^{2,3}$,  \\ 
Roman Stocker$^{2,3}$ \& Simon A. Levin$^{1}$
\\
\normalsize{$^{1}$Department of Ecology and Evolutionary Biology, Princeton University,} \\
\normalsize{Princeton, NJ 08544, USA}\\
\normalsize{$^{2}$Ralph M. Parsons Laboratory, Department of Civil and Environmental Engineering,}\\  \normalsize{Massachusetts Institute of Technology, Cambridge, MA 02139, USA}\\
\normalsize{$^{3}$Department of Civil, Environmental and Geomatic Engineering,}\\  \normalsize{ETH Zurich, 8093 Zurich, Switzerland}\\
\normalsize{$^*$ Email for correspondence: ahein@princeton.edu}
\\
}

\date{}

\begin{document} 
\baselineskip18pt
\maketitle 

\abstract{Many chemotactic bacteria inhabit environments in which chemicals appear as localized pulses and evolve by processes such as diffusion and mixing. We show that, in such environments, physical limits on the accuracy of temporal gradient sensing govern when and where bacteria can accurately measure the cues they use to navigate. Chemical pulses are surrounded by a predictable dynamic region, outside which bacterial cells cannot resolve gradients above noise. The outer boundary of this region initially expands in proportion to $\sqrt{t}$, before rapidly contracting. Our analysis also reveals how chemokinesis -- the increase in swimming speed many bacteria exhibit when absolute chemical concentration exceeds a threshold -- may serve to enhance chemotactic accuracy and sensitivity when the chemical landscape is dynamic. More generally, our framework provides a rigorous method for partitioning bacteria into populations that are ``near" and ``far" from chemical hotspots in complex, rapidly evolving environments such as those that dominate aquatic ecosystems.}
\newline


\section*{Introduction}

In natural environments such as oceans and lakes, bacteria and other microbes navigate chemical landscapes that can change dramatically over the timescales relevant to their motility \cite{taylor:2012}. Such environments differ in fundamental ways from the static chemical gradients typically considered in studies of microbial chemotaxis (e.g.,\cite{kalinin:2009,ahmed:2010}). From the perspective of microbes, chemical cues in nature often appear as localized pulses with short duration \cite{stocker:2012,blackburn:1998}. For example, oil droplets from spills and natural seeps, organic matter exuded by lysed phytoplankton or excreted by other organisms, and marine particles are common sources of short-lived, micro-scale ($\sim$10-1000 $\mu$m) chemical pulses \cite{stocker:2012}. Motile bacteria respond to such cues by swimming up the gradients that are generated when pulses diffuse (e.g., \cite{blackburn:1998, seymour:2009, stocker:2008, seymour:2010}). When a pulse appears, for example through the lysis of a phytoplankton cell, the distribution of chemoattractants (often, dissolved organic matter) changes rapidly over both space and time \cite{fenchel:2002}. Because background conditions are highly dilute, bacteria experience the early stages of a spreading pulse as a noisy chemical gradient with low absolute concentration. In marine environments, ephemeral, micro-scale pulses of dissolved chemicals provide a substantial and perhaps dominant fraction of the resources used by heterotrophic bacteria \cite{stocker:2012,fenchel:2002,barbara:2003}. {The advantage that chemotaxis confers cells in such dynamic environments \cite{frankel:2014,celani:2010,taylor:2012} may help explain why chemotactic responses to transient nutrient sources are so common among marine bacteria \cite{barbara:2003, blackburn:1998,seymour:2009,seymour:2010}. }

{Although chemotaxis appears to be an important driver of bacterial competition \cite{taylor:2012}, evolution \cite{celani:2010,frankel:2014}, and nutrient cycling \cite{stocker:2012,fenchel:2002}, the details of bacterial chemotaxis behaviour are poorly characterized for all but a few well-studied species of bacteria. An important shared feature of bacterial chemotaxis systems, however, is that the measurements of chemical concentration that underpin chemotaxis behavior are subject to considerable noise \cite{mora:2010,andrews:2006}. In particular, stochasticity in the times at which individual molecules of chemoattractant arrive at the bacterium's surface sets an upper bound on the precision with which the cell can measure changes in concentration \cite{berg:1977,bialek:2005}. Here, we demonstrate how this physical limit on the precision of temporal gradient sensing constrains when and where bacteria can respond to chemical pulses. Using this approach, we develop a general theory to predict the fundamental length and timescales over which chemotactic bacteria can respond to chemical pulses. Because it requires few assumptions about the underlying mechanisms responsible for chemotactic behaviour, the theory can be applied to the diverse assemblages of bacteria that occur in natural marine and freshwater environments.} 

We first discuss gradient estimation by a cell in a dynamic chemoattractant field. We then derive theoretical bounds on the regions of the environment in which bacteria can respond to gradients, and characterize the spatio-temporal evolution of these regions as a function of physical and biological parameters. Finally, we show that changes in swimming speed in response to measurements of absolute concentration -- a bacterial behaviour known as chemokinesis \cite{barbara:2003, garren:2013} -- can greatly enhance a cell's ability to measure gradients in a dynamic chemoattractant field.

\section*{Model development}
\subsection*{Signal and noise in temporal gradient sensing} 
\begin{figure}[h]
\captionsetup{font=small,width=16cm}
\begin{center}
\includegraphics[width=9.2cm]{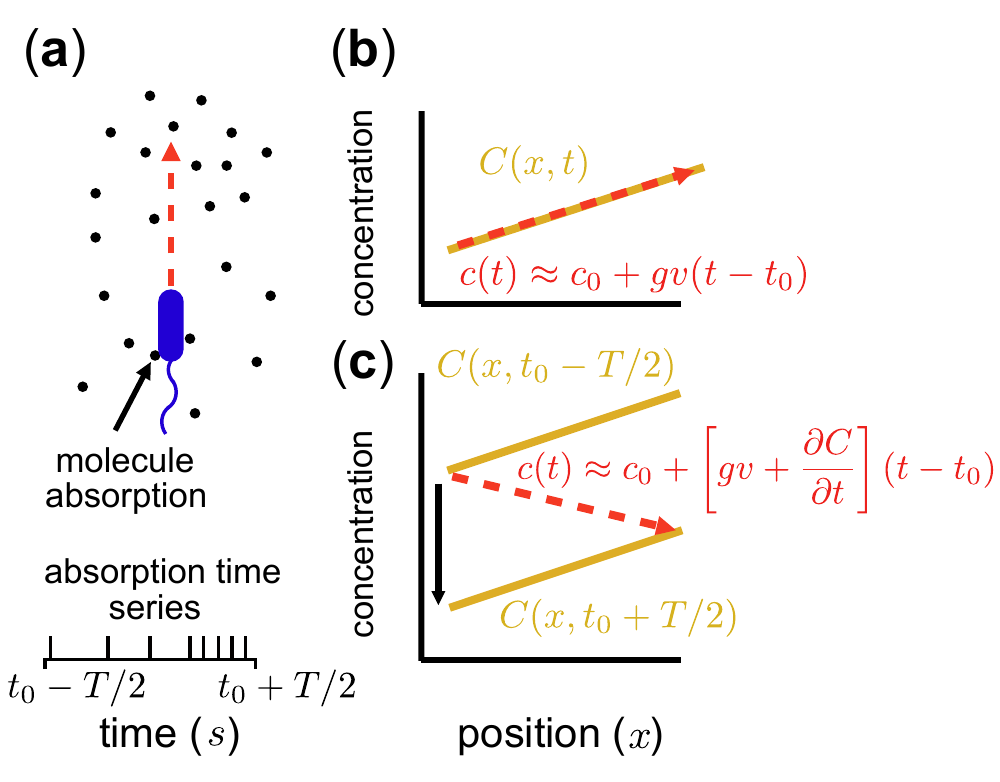}
\end{center}
\caption{Measurement of ramp rate $c_1$ by an idealized cell. (a) During a time interval of length $T$, a cell travels from a region of low concentration to a region of higher concentration, absorbing chemoattractant molecules at times $\{t_i \}$ (red spikes in time series). (b) In a static concentration field $C(x)$, $c_1$ is equal to concentration slope $g$ (slope of orange line) times swimming speed $v$. (c) In a dynamic concentration field $C(x,t)$, $c_1 \approx vg + \partial C/\partial t$; $g$ is confounded with temporal changes in concentration ($\partial C/\partial t$) and the cell may perceive a decreasing concentration (red dashed line) although the true concentration slope is positive. Figures in colour online.}
\label{fig1}
\end{figure}

Unlike large eukaryotic cells, which can directly measure spatial gradients in chemical concentration \cite{endres:2008}, many chemotactic bacteria navigate by measuring temporal changes in concentration as they swim \cite{macnab:1972, segall:1986}. They use these measurements to detect concentration gradients and to navigate toward more favourable conditions (toward resources, away from noxious substances). Regardless of the biochemical and behavioural mechanisms a cell uses to navigate, gradient-based navigation can only be as precise as a cell's estimate of the gradient itself; downstream transduction will, in general, only add noise \cite{bialek:2005}. One can, therefore, establish performance bounds within which real bacterial cells must operate by considering physical limits on the accuracy and precision of gradient sensing by an idealized cell. We begin by considering gradient detection by such a cell: the perfectly absorbing sphere originally described by Berg and Purcell \cite{berg:1977}. This cell swims through a dynamic chemoattractant landscape, absorbing all molecules that reach its surface (Fig.~\ref{fig1}a). In reality, bacteria absorb some ligands they use for chemotaxis, whereas others are bound only temporarily. However, absorbing ligand always leads to more accurate measurement of both absolute concentration and changes in concentration over time because molecules cannot be re-bound once they have been absorbed \cite{endres:2008, mora:2010}. We therefore assume molecules are absorbed yielding an upper limit on measurement accuracy \cite{endres:2008}.

Like the well-studied enteric bacterium, \emph{Escherichia coli}, marine bacteria perform chemotaxis by altering the length of relatively straight ``runs", which are interspersed with random re-orientation events (``tumbles" for \textit{E. coli} \cite{brown:1974}, ``flicks" for marine bacteria \cite{xie:2011,xie:2015}). As a cell swims, receptors on the cell's surface bind chemoattractant molecules and a signal from the receptors is transduced through a biochemical network to one or more flagellar motors, which control the speed and direction of the flagellar rotations that drive locomotion. Changes in receptor occupancy alter the probability that the direction of flagellar rotation will reverse, leading to a re-orientation \cite{jiang:2010}, and the outcome of this is that bacteria extend runs when they perceive an increasing concentration of chemoattractant. A requirement for chemotaxis, therefore, is that the cell is capable of detecting meaningful changes in mean concentration \cite{andrews:2006} over some measurement interval of length $T$. This task is complicated by significant stochastic variation in the times at which molecules arrive at the cell's surface. The length of the measurement interval $T$ is bounded above by the characteristic timescale of stochastic re-orientations (e.g., rotational diffusion, active re-orientation \cite{berg:1977}), which for cells in the size range of \emph{E. coli} and many marine bacteria, ranges from hundreds of milliseconds \cite{blackburn:1998} to several seconds \cite{alon:1998}. A cell has little to gain by using the history of molecule encounters that extends beyond this timescale because rotational diffusion and active stochastic reorientation (e.g., tumbles, flicks) cause random changes in the cell's trajectory, decorrelating the cell's orientation, and rendering old information useless to the cell for determining whether it is currently travelling up or down a chemoattractant gradient (this issue is discussed in detail in \cite{berg:1977}). We therefore assume the measurement timescale $T$ is shorter than the timescale of stochastic reorientation and neglect processes such as rotational diffusion. For such short $T$, the chemoattractant concentration along the swimming cell's path, $c(t)$, can be linearized to $c(t) \approx c_0  + c_1 (t-t_0)$ over the time interval $(t_0-T/2,t_0+T/2)$. The cell experiences this concentration as a noisy time series of encounters with chemoattractant molecules (Fig.~\ref{fig1}a), from which it must estimate the concentration ramp rate, $c_1$, to determine whether concentration is increasing or decreasing.

Using maximum likelihood, one can show that the optimal way for a perfectly absorbing sphere of radius $a$ to estimate $c_1$ (concentration $\times$ time$^{-1}$) using a sequence of molecule absorptions is, to leading order \cite{mora:2010}: $\hat{c_1} = \frac{n\sum_i (t_i - t_0)}{4\pi D a T\sum_i (t_i - t_0)^2}$, where $\hat{c_1}$  is the cell's estimate of the ramp rate, $n$ is the number of molecules absorbed over the measurement interval, $D$  is the diffusivity of the chemoattractant, and $t_i$ is the absorption time of the $i$th molecule. Importantly, $\hat{c_1}$ has typical measurement variance no less than:

\begin{equation}
Var(\hat{c_1}) = \frac{ 3 c_0 }{\pi D a T^3},
\label{eq:var}
\end{equation}
where $c_0$ is the true background concentration in the vicinity of the cell at time $t_0$, and the variance of $\hat{c_1}$ does not depend on the true ramp rate $c_1$ as long as $c_0 \gg c_1 T$ (Supplementary Text, {see also Equation (S44) in ref. \cite{mora:2010}}). This formulation assumes that a cell can ``count" many molecules in a typical observation window, which amounts to assuming that the timescale at which receptors bind chemoattractant molecules is fast relative to the length of the observation window, $T$. Receptor binding kinetics are typically {very} fast (millisecond timescales, e.g. \cite{zhang:2005, jiang:2010}), so this assumption will generally hold unless $T$ is {extremely} short. {To summarize, measurements of concentration involve three timescales that are relevant to our model formulation, which are naturally separated in chemotactic bacteria \cite{jiang:2010}: (1) the timescale of absorptions, which is typically short ($\sim$1 ms \cite{jiang:2010}), (2) the measurement window $T$, which is of intermediate length, and (3) the timescale of active re-orientations, which must be longer than $T$ if the bacterium is to perform chemotaxis \cite{berg:1977}.}

Variance in the ramp rate estimate  (Eq.~(\ref{eq:var})) is solely due to stochastic arrivals of chemoattractant molecules and does not include additional sources of noise resulting, for example, from noise in the biochemical network responsible for ramp rate estimation \cite{bialek:2005,lestas:2010}. Eq.~(\ref{eq:var}) thus provides a lower bound on uncertainty about the true ramp rate and a constraint within which real cells must operate, regardless of the precise biochemical mechanism though which they implement ramp rate estimation. Below we use Eq.~(\ref{eq:var}) to define the regions of space where it is possible for cells to use measurements of concentration to climb chemoattractant gradients. Outside these regions, cells may attempt to perform chemotaxis; however, we will show that for several ecologically relevant types of pulses, the signal-to-noise ratio of a cell's estimate of the concentration slope decays sharply (like a Gaussian) far from the origin of a chemoattractant pulse. This strong decrease in the signal-to-noise ratio with increasing distance implies that chemotactic cells far from the origin of a pulse will be responding primarily to noise and will not exhibit biased motion.

\subsection*{Gradient estimation in a time-varying environment}  For a cell swimming at speed $v$, the instantaneous local slope of the concentration profile along the cell's path, which we will refer to as the concentration slope $g$, is given by $g = \nabla C(\mathbf{x}) \cdot \mathbf{v}/v$, where $\mathbf{v}$ is the cell's velocity. The concentration slope is the quantity that is useful for climbing gradients, for example, by providing a signal for cells to lengthen runs in run-and-tumble chemotaxis \cite{brown:1974}; however, a cell the size of a bacterium ($\sim$1 $\mu$m) cannot measure $g$ directly \cite{mora:2010}. It must instead infer $g$ from its estimate of the ramp rate $\hat{c_1}$. In a time-invariant concentration field $c_1=gv$, and the maximum likelihood estimator of $g$ is proportional to the ramp rate estimator: $\hat{g}=\hat{c_1}/v$ (Fig.~1b, Supplementary Text). 

In a time-varying environment the concentration that a swimming cell experiences, $c(t) \approx c_0  + (v g + \partial C/\partial t) (t-t_0)$, is influenced by local temporal changes in concentration, $\partial C/\partial t$ (Fig.~1c); the ramp rate is given by $c_1=vg + \partial C/\partial t$. In this case, the time series of molecule absorptions does not contain the information needed to estimate both $g$ and $\partial C/\partial t$ and any estimator the cell uses to measure the concentration slope $g$ will be biased (Supplementary Text). For example, estimating $g$ as $\hat{g} = \hat{c_1}/v $ means that $\hat{g} \rightarrow g + (\partial C / \partial t) /v$ in the limit of many molecule absorptions. Correcting this bias would require that the cell have an independent estimate of $\partial C / \partial t$. In the absence of such an estimate, the cell can reduce bias by travelling faster, but not by increasing the length of its measurement window $T$ (Supplementary Text). This highlights an important connection between swimming speed and measurement accuracy that we explore in more detail below. Bias in the concentration slope estimate becomes important far from the origin of a pulse where cells can perceive an increasing concentration even if they are travelling down the concentration gradient, and near the origin, where cells can perceive a falling concentration even if they are travelling up a gradient (Fig.~\ref{fig1}c).

\subsection*{Conditions for chemotaxis and responses to chemical pulses} If a cell is to use measurements of ramp rate to climb a concentration gradient, two conditions must be met. First, the cell must be in a region of the environment where typical values of the perceived ramp rate exceed noise: i.e., the signal-to-noise ratio (SNR) of the ramp rate estimator, $|c_1|Var(\hat{c_1})^{-1/2} \geq \delta_0$, where $\delta_0$ is a constant threshold on the SNR (Supplementary Text). Second, the ramp rate $c_1=vg + \partial C/\partial t$ and the concentration slope $g$ must have the same sign. Applying Eq.~(\ref{eq:var}) and rearranging, these conditions are:
\begin{gather}
\frac{|vg + \frac{\partial C }{\partial t}|}{\sqrt{c_0}}  \geq \delta := \delta_0\sqrt{\frac{3}{\pi D a T^3}} ,\nonumber \\
 \mathrm{and}\label{eq:conditions}\\
  \; \mathrm{sign}(c_1) = \mathrm{sign}(g). \nonumber 
\end{gather}
For a chemoattractant field with concentration $C(\mathbf{x},t)$, Conditions (\ref{eq:conditions}) define the regions where cells can reliably determine the sign of the concentration slope, a requirement for gradient-based navigation.

Using Conditions (\ref{eq:conditions}), we explore how bacteria perceive three types of pulses that occur in natural environments: pulses that arise from surfaces, pulses that arise as thin chemical filaments, and pulses created by small point releases. Localized point pulses are created by many natural sources, including the lysis of small cells and excretions by larger organisms \cite{stocker:2012,blackburn:1998}. Thin chemical filaments and sheets occur when turbulence stirs dissolved chemicals. The distribution of chemicals is stretched and folded into sheets and filaments at length scales down to the Batchelor scale \cite{stocker:2012}. Mixing below the Batchelor scale is dominated by diffusion. This length scale is $l_B = (\nu D^2/ \epsilon)^{1/4}$, where $\nu$ is kinematic viscocity, $D$ is mass diffusivity, and $\epsilon$ is the turbulent dissipation rate. As $\epsilon$ changes, $l_B$ changes slowly implying that small point pulses and filaments or sheets spread primarily by diffusion across a broad range of flows. {Across a range of realistic levels of turbulence ($\epsilon \sim 10^{-9}$ to $10^{-6}$ W kg$^{-1}$ \cite{doubell:2014}) the average shear rate is of order $10^{-3}$ to 1 s$^{-1}$. Except for the highest values in this range, these shear rates are typically too low to cause significant re-orientation of bacteria as they swim \cite{rusconi:2014}. We therefore focus on the regime in which the environment is steady over the length scales considered here.}  

{To illustrate the utility of our theory, we consider how bacteria respond to chemical point pulses, filaments, and sheets.} These canonical geometries can be viewed as basic components of more complex chemical landscapes at larger scales (e.g., the types of landscapes considered in \cite{taylor:2012}). Extending our results to alternative geometries follows from straightforward calculations. At time, $t = 0$, a single pulse appears with planar ($N$ = 1, sheet), cylindrical ($N$ = 2, filament), or spherical ($N$ = 3, point pulse) symmetry. The size of the pulse is $M$ (molecules per unit area of sheet [$N$ = 1], per unit filament length [$N$ = 2], or per individual point pulse [$N$ = 3]). The three-dimensional chemoattractant field $C$ is governed by $\partial C /\partial t = D \Delta C$ and the concentration is:

\begin{equation}
C(r, t, N) = \frac{M}{(4\pi D t)^{N/2}}e^{-\frac{r^2}{4Dt}},
\label{conc_radial}
\end{equation}
where $D$ ($\mu$m$^2$ s$^{-1}$) is diffusivity, $r$ ($\mu$m) is the distance from the surface ($N = 1$), filament axis ($N = 2$), or centre of the point source ($N = 3$). A cell moving in this chemoattractant field with velocity $v$ ($\mu$m$\,$s$^ {-1}$) will experience a typical rate of change in concentration of $c_1\approx \nabla C \cdot \mathbf{v} + \partial C / \partial t$. 

\begin{figure}[t!]
\captionsetup{font=small,width=16cm}
\begin{center} 
\includegraphics[width=12cm]{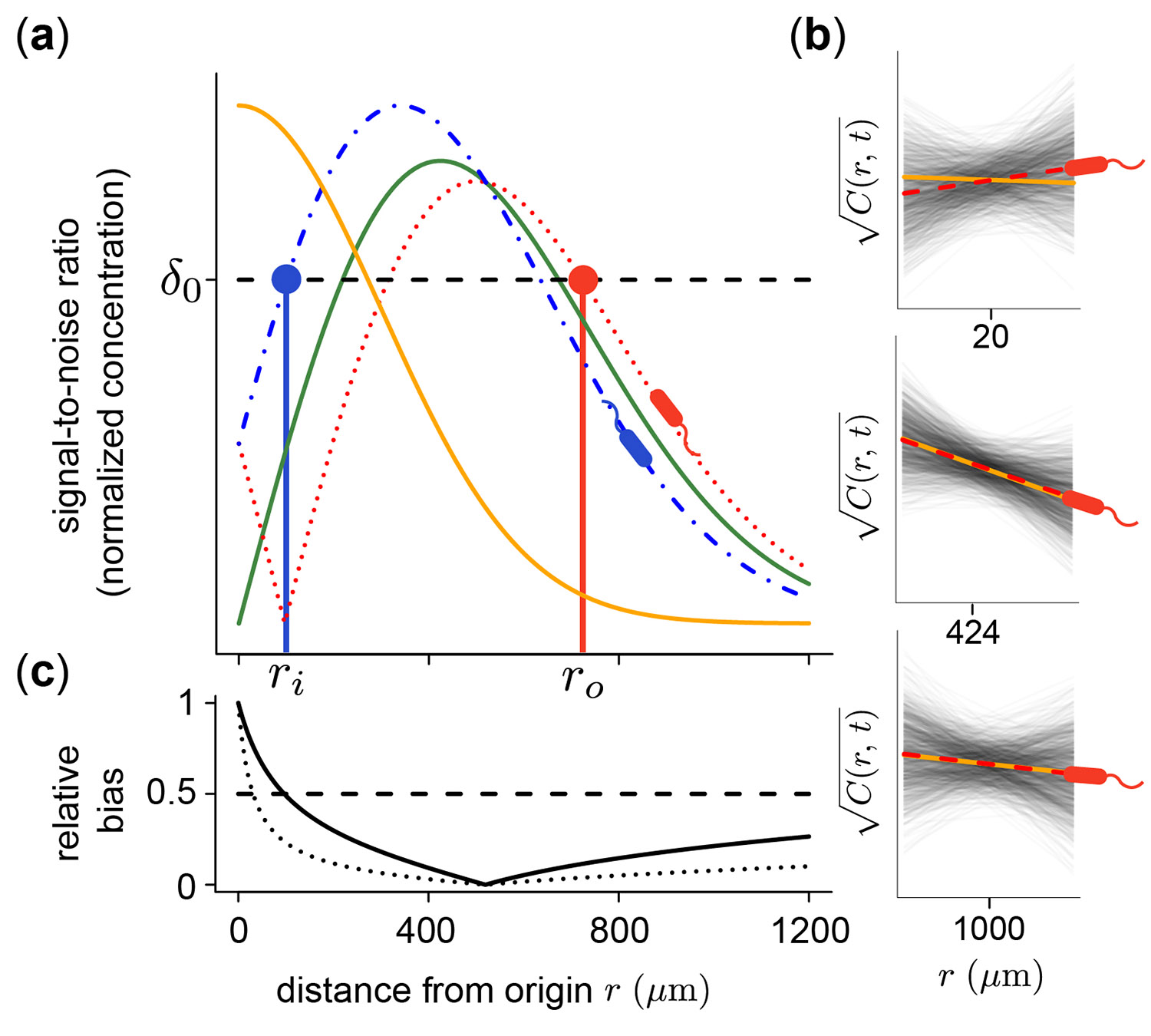}
\end{center}
\caption{Gradient estimation in a dynamic environment. (a) Solid orange curve shows the true concentration profile at $t = t_0$. Solid green curve shows the signal-to-noise ratio (SNR) of $\hat{c_1}$ a cell would experience if this concentration profile were static. Dotted red curve shows SNR for a cell swimming directly toward origin of pulse. Dashed blue curve shows SNR for a cell swimming directly away from origin of pulse. Concentration and SNR normalized to maximum value of one. {(b) Square-root of concentration ($\sqrt{C(r,t)}$) at $t = t_0$ (orange) and individual estimates of this concentration ($\sqrt{c(t)}$, grey) made by a cell swimming toward pulse origin. Each individual estimate is computed by calculating $\hat{c_0}$ and $\hat{c_1}$ (see Supplementary Text for equations) from a time series of random Poisson molecule arrivals \cite{asmussen:2007} with an arrival rate given by the true instantaneous concentration at the bacterium's position $C(\mathbf{x},t)$.} (c) Relative bias of concentration slope estimate ($ |\partial C /\partial t | / [ |vg| + |\partial C/\partial t | ]$) measured by slow (solid curve; $v = 30\, ~\mu$m$\, \text{s}^{-1}$) and fast swimming cells (dotted curve; $v = 96$ $\mu$m$\,\text{s}^{-1}$). In all panels, concentration governed by Eq.~(\ref{conc_radial}) with $N = 3$, $M = 10^{11}$ molecules, $v = 30\, \mu$m s$^{-1}$, $a = 1 \, \mu$m, $T = 0.1 \,$s, $t_0 = 45 \,$s, and $\delta_0 = 1$. Pulse sizes in all figures correspond roughly to the quantity of free amino acids released from a lysed phytoplankton cell of $\sim 10$ $\mu$m in diameter \cite{blackburn:1998}. }
\label{noisy_grad}
\end{figure}

For chemoattractant pulses with concentration described by Eq.~(\ref{conc_radial}) (Fig.~\ref{noisy_grad}a, solid orange curve), the signal-to-noise ratio (Fig.~\ref{noisy_grad}a, solid green curve) divides the domain surrounding a pulse into three regions. Far from the pulse, the concentration gradient is shallow and the absolute concentration is low: cells cannot accurately measure changes in concentration because they encounter few molecules during a typical observation window (Fig.~\ref{noisy_grad}b, bottom panel). At an intermediate distance from the pulse origin, the gradient is largest in magnitude and cells encounter many molecules during a typical observation window: the SNR is greatest in this region (Fig.~\ref{noisy_grad}b, middle panel). Near the pulse origin the gradient is again shallow and variance in the concentration slope estimate is substantial (Fig.~\ref{noisy_grad}b, top panel). Moreover, in this region, concentration changes rapidly over time and the concentration slope and ramp rate may differ in sign (i.e., bias in the concentration slope estimate is large, Fig.~\ref{noisy_grad}b, top panel; Fig.~\ref{noisy_grad}c). 

\section*{Results}
Cells far from a chemoattractant pulse cannot resolve true changes in concentration above noise (Fig.~\ref{noisy_grad}a, SNR drops below threshold $\delta_0$ for large distance). The distance beyond which $\hat{c_1}$ becomes dominated by noise is given implicitly by
\begin{equation}
\delta= \left| vg(r,t) + \frac{\partial C(r,t)}{ \partial t} \right| C(r,t)^{-1/2},
\label{del}
\end{equation}
where the term in brackets is the magnitude of the true ramp rate $c_1$ that a cell at distance $r$ with local concentration slope $g(r,t)$ experiences. Because the chemoattractant field is changing, the magnitude of the ramp rate a cell measures will depend on its direction of travel. Far from the pulse, a cell travelling directly inward (Fig.~\ref{noisy_grad}a, red dotted curve) will experience a greater SNR than a cell travelling outward (Fig.~\ref{noisy_grad}a, blue dot-dash curve). Beyond the inflection point in the concentration profile, the r.h.s. of Eq.~(\ref{del}) is maximized for cells travelling directly up the concentration gradient (i.e., toward the pulse center; Fig.~\ref{noisy_grad}a, red dotted curve). The outer boundary beyond which cells cannot reliably perceive changes in concentration is given implicitly by Eq.~(\ref{del}) with $g= -\partial C/\partial r$. We refer to the largest distance that satisfies this equation as the outer boundary of sensitivity, $r_o$ (Fig.~\ref{noisy_grad}a, red point). At distances $r>r_o$, perceived changes in concentration are dominated by noise, regardless of a cell's direction of travel.  

Bacteria use gradients to navigate toward regions of high attractant concentration, but also to maintain position near local maxima \cite{celani:2010}. In order to do this, a cell travelling down the concentration gradient must experience a decreasing concentration, which provides the signal the cell uses to modify swimming behavior \cite{xie:2015}. Near the origin, the SNR is maximized for a cell that is travelling directly down the concentration gradient (Fig.~2a blue dash-dot curve). For $t$ greater than a critical time, $t_s$, there is an inner boundary at a distance $r_i$ from the origin of the pulse (Fig.~2a, blue point), within which the SNR drops below threshold. For $t>t_s$ the location of this inner boundary is given implicitly by Eq.~(\ref{del}) with $g= \partial C(r,t)/ \partial r$ (Supplementary Text). 

The boundaries $r_o$ and $r_i$ define a dynamic region (Fig.~\ref{fig:inner_outer}, blue region in inset), outside of which bacteria cannot reliably respond to chemoattractant gradients because either the ramp rate is too noisy to resolve, or the ramp rate and the concentration slope have different signs (i.e., Conditions (\ref{eq:conditions}) are violated).  Figure \ref{fig:inner_outer} shows the dynamics of $r_o$ and $r_i$ for bacteria swimming at three different speeds. For all swimming speeds, the outer boundary $r_o$ initially expands before rapidly contracting (Fig.~\ref{fig:inner_outer}, red curves). The time dependence of this boundary can be obtained by substituting Eq.~(\ref{conc_radial}) into Eq.~(\ref{del}), solving for $r_o$, and expanding the resulting product-log solution (Supplementary Text):

\begin{figure}[t!]
\captionsetup{font=small,width=16cm}
\begin{center}
\includegraphics[width=9.3cm]{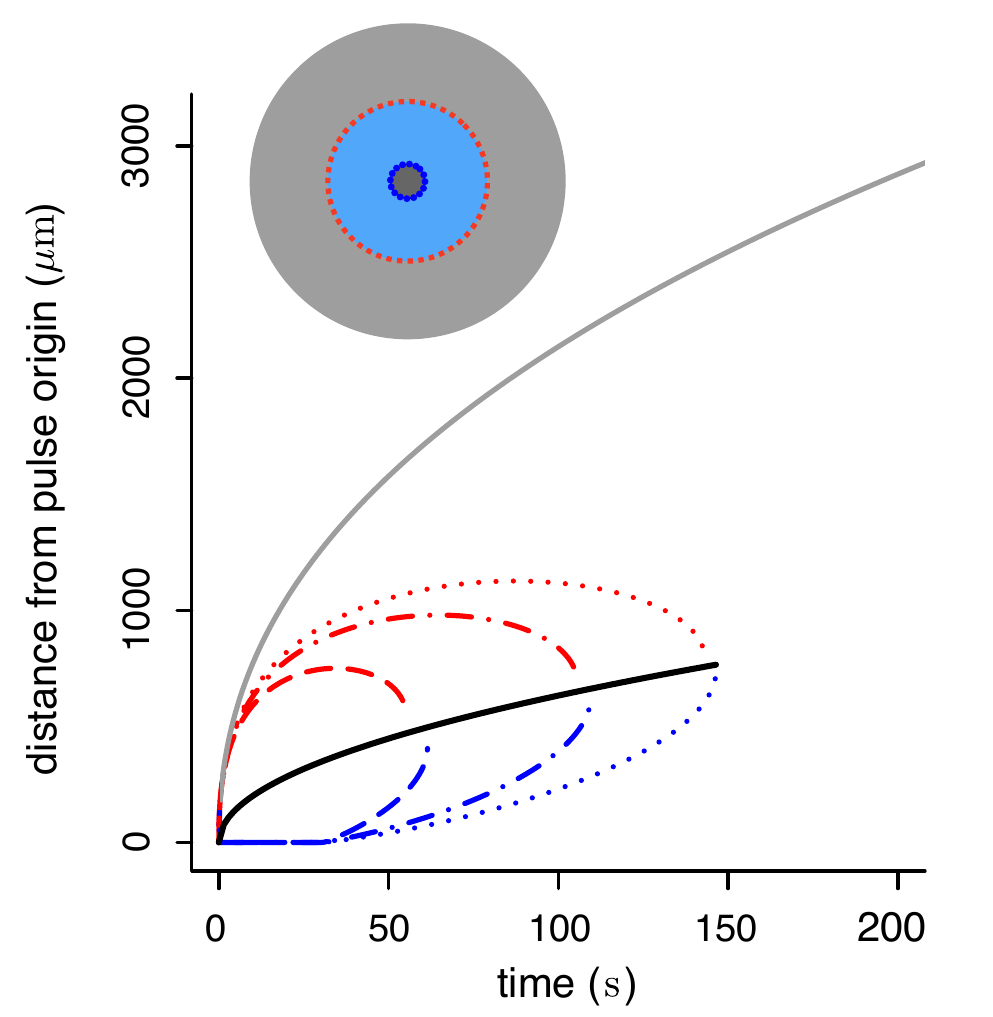}
\end{center}
\caption{Inner (blue) and outer (red) boundaries of the region in which cells reliably perceive gradients. Dashed line shows $v = 30\, \mu$m$\,$s$^{-1}$, maximum swimming speeds of \emph{E. coli} \cite{barbara:2003}; dash-dot line shows $v = 66\, \mu \text{m s}^{-1}$, typical cruising speed of \emph{Vibrio coralliilyticus}; dotted line shows $v = 96 \, \mu \text{m s}^{-1}$, maximum speed of \emph{V. coralliilyticus} after initiating chemokinesis \cite{garren:2013}. Other parameters as in Fig.~2. Solid grey curve is outer boundary, $r_c$, of region within which cells can resolve absolute concentration. Solid black curve is $\sqrt{4Dt}$, the radius at which the SNR is maximized for a static profile (green curve in Fig.~2). Inset shows relative sizes of region where cells can detect gradients ($r_i < r <r_o$, blue region), and region where cells can resolve absolute concentration ($r < r_c$, grey region inward) at $t = 90\,$s ($v = 66$ $\mu$m$\,$s$^{-1}$).}
\label{fig:inner_outer}
\end{figure}

\begin{equation}
r_o \approx \sqrt{4 D t \log \left[ \frac{-\log  (kt^{1 + N/2})}{k t^{1 + N/2}} \right] },
\label{roapprox}
\end{equation}
where $k =  (4 \pi D)^{N/2} \delta_0^2 /(2 \pi a M v^2 T^3)$. Swimming speeds of motile bacteria typically range from $30 \, \mu$m$\,$s$^{-1}$ to over $100\, \mu$m$\,$s$^{-1}$\cite{barbara:2003}. For many relevant chemoattractants, $D \sim 10^3\, \mu$m$^2$ s$^{-1}$, and the number of molecules released in a pulse, $M$, is generally large; for example, a point pulse created by the lysis of even a small phytoplankton cell (a common source of nutrients for marine bacteria) contains upwards of $10^{11}$ free amino acid molecules \cite{blackburn:1998}. This means that $k \ll 1$ such that the logarithmic term in Eq.~\eqref{roapprox} varies slowly with time for early times, and leading-order behaviour is initially governed by $\sqrt{t}$. Pulse size, $M$ occurs only inside the logarithmic terms in Eq.~(\ref{roapprox}) indicating that $r_o$ scales weakly with pulse size. For example, doubling the size of a small point pulse ($N = 3$) increases the volume of water in which gradients are perceived by only 50\% (assuming $M$ increases from $10^{11}$ to $2 \times 10^{11}$ molecules, $\delta_0 = 1$, and $v=66$ $\mu$m s$^{-1}$). Figure \ref{fig:ro_approx} shows the dynamics of $r_o$ for surface, filament, and point pulses. Eq.~(\ref{roapprox}) agrees well with the exact solution for $r_o$ obtained by solving Eq.~(\ref{del}) numerically (Fig.~\ref{fig:ro_approx} compare solid and dashed lines). 

Eventually the inner and outer boundaries of sensitivity intersect (Fig.~\ref{fig:inner_outer}), and cells can no longer reliably glean navigational information from the chemoattractant field. We refer to the time at which this occurs as $t^*$. Finding the time when the SNR falls below threshold $\delta_0$ everywhere shows that

\begin{equation}
t^* \approx \alpha(Mv^2T^3)^{\frac{2}{N+2}},
\label{tstar}
\end{equation}
where $\alpha = (\pi^{(1-N/2)} a e^{-1})^{2/(N+2)} [3(4D )^{N/2} \delta_0]^{-2/(N+2)}$ and the approximation assumes $|vg| \gg \partial C/\partial t$ at the point in space where the SNR is maximized (Supplementary Text). This relation illustrates the relative contribution of measurement time $T$ and speed $v$ to the time scale of perceptible changes in concentration, $t^*$. Moreover, Eq.~\eqref{tstar} shows that $t^*$ is proportional to $M^{2/(N + 2)}$; the scaling of $t^*$ with pulse size is sublinear for all pulse geometries meaning that doubling the size of a pulse always less than doubles the time over which it can be perceived. 

\begin{figure}[t!]
\captionsetup{font=small,width=16cm}
\begin{center}
\includegraphics[width=11cm]{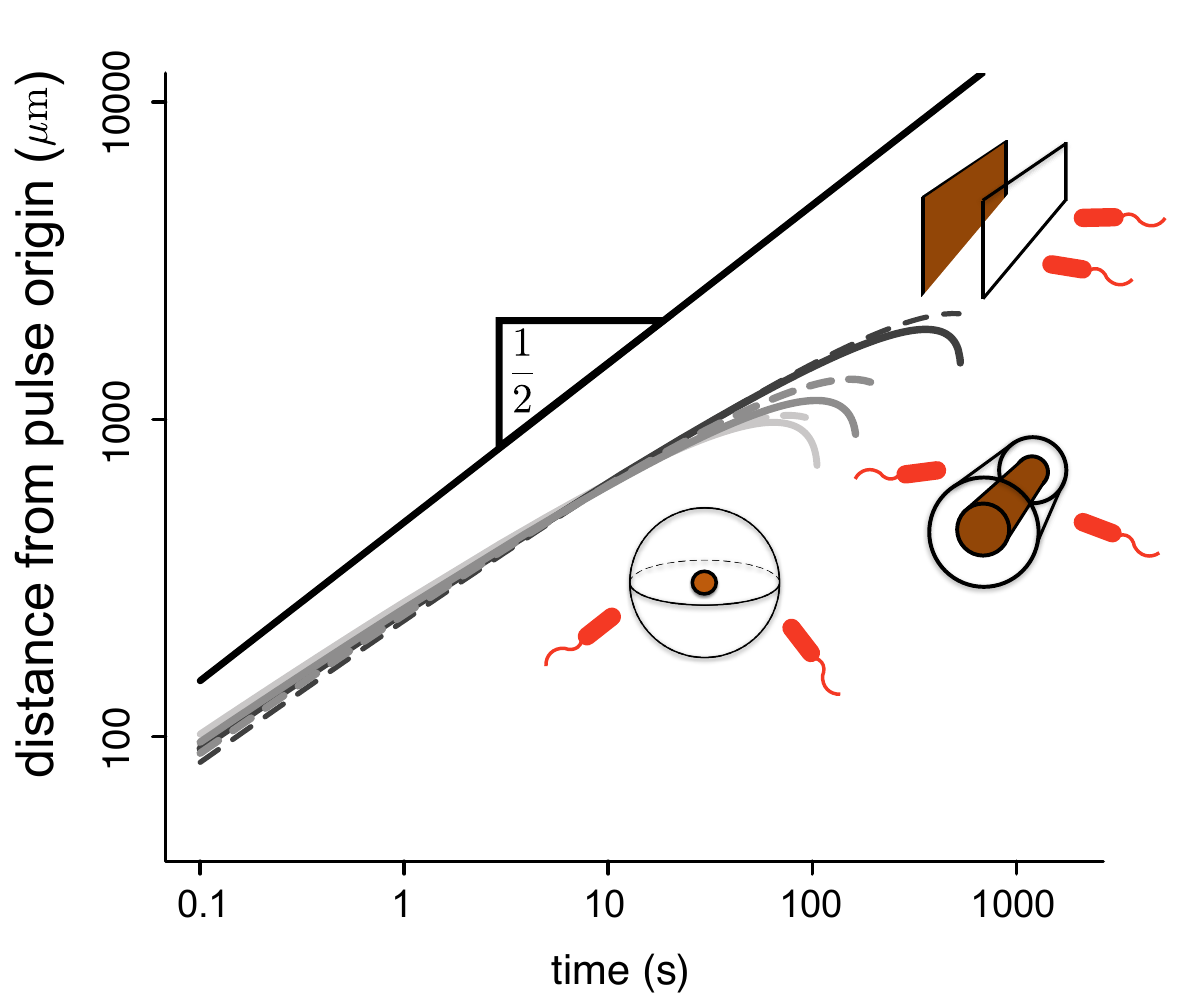}
\end{center}
\caption{Scaling of the outer boundary of sensitivity $r_o$ for pulses emitted from surfaces (light grey), filaments (grey), and point sources (black). Solid curves are numerical solution to Eq.~(\ref{del}). Dashed curves given by Eq.~(\ref{roapprox}). Solid black line is proportional to $\sqrt{t}$. Solid curves truncated when the SNR falls below $\delta_0$. Dashed curves truncated at $t^*$ (Eq.~(\ref{tstar})). $M$ scaled so that pulses with different geometries have the same concentration profile at $t = 10$ s ($M = 8.0 \times 10^5$ molecules per $\mu$m$^2$ surface for surface source; $M = 2.8 \times 10^8$ molecules per $\mu$m length for line source; $M = 10^{11}$  molecules for point source); $v = 66\, \mu$m s$^{-1}$; other parameters as in Fig.~2.}
\label{fig:ro_approx}
\end{figure}

The locations of inner and outer boundaries (Fig.~\ref{fig:inner_outer}) are governed, in part, by swimming speed. {Many bacteria alter swimming speed in response to stimuli, and} a natural question, therefore, is whether a cell could adjust its speed adaptively to achieve high sensitivity {to chemical gradients.} Some species exhibit a behaviour known as chemokinesis: cells swim at a speed that depends on the local concentration of chemoattractant, often swimming at a high speed when absolute concentration is high, and a low speed when concentration is low \cite{barbara:2003,garren:2013}. In the presence of a resolvable gradient, the interpretation of chemokinesis is straightforward: cells can climb the gradient faster if they swim at a higher speed (at the expense of a higher energetic cost of motility). However, chemokinesis may also have a second role. The SNR of the ramp rate is smaller than the SNR of the absolute concentration, $c_0$, implying that cells may be able to accurately detect whether absolute concentration has crossed a threshold before they can resolve changes in concentration over time. The mean rate of arrival of molecules to the surface of a sphere of radius $a$ is $4\pi D ac(t)$ \cite{berg:1977}. Poisson molecule arrivals imply that the SNR of absolute concentration $c_0$ is $c_0 Var (\hat{c_0} )^{-1/2} = c_0 [4\pi D aTc_0]^{-1/2}$. Using this ratio, we define a third boundary, $r_c$, beyond which the SNR of $\hat{c_0}$ falls below threshold, $\delta_0$:
\begin{equation}
r_c = \sqrt{8Dt \log(\eta t^{-N/2})},
\label{eq:rc}
\end{equation}
where $\eta = \delta_0^{-1} (MaT)^{1/2} (4\pi D)^{1/2-N/4}$. This boundary has the same leading order behaviour in time as $r_o$, but extends well beyond $r_o$ (Fig.~3, solid grey curve); for example, assuming $r_o$ is at its maximum value (Fig.~3), the volume within which cells can accurately measure absolute concentration in the water surrounding a small point pulse ($N = 3$) is six times larger than the volume in which cells can resolve changes in concentration (assuming $M = 10^{11}$ molecules \cite{blackburn:1998}, $\delta_0 = 1$, $v = 66$ $\mu$m s$^{-1}$).  Note that we use the same threshold ($\delta_0$) on the SNR of $\hat{c_0}$ and $\hat{c_1}$ for the purpose of comparison but thresholds on these ratios need not be equal.

By increasing their swimming speeds when concentration exceeds a threshold, cells can increase their sensitivity to changes in concentration (first Condition (\ref{eq:conditions}); Fig.~\ref{fig:inner_outer}) and reduce bias in estimation of the concentration slope (Fig.~2c). The effect of increasing swimming speed is to expand the region of space over which the cell can resolve gradients, $r_i < r < r_o$, and to extend the time $t^*$ beyond which gradients become too noisy for the cell to measure (Fig.~\ref{fig:inner_outer}, compare curves for different swimming speeds; Fig.~5). 

\begin{figure}[t!]
\captionsetup{font=small,width=16cm}
\begin{center}
\includegraphics[width=11.5cm]{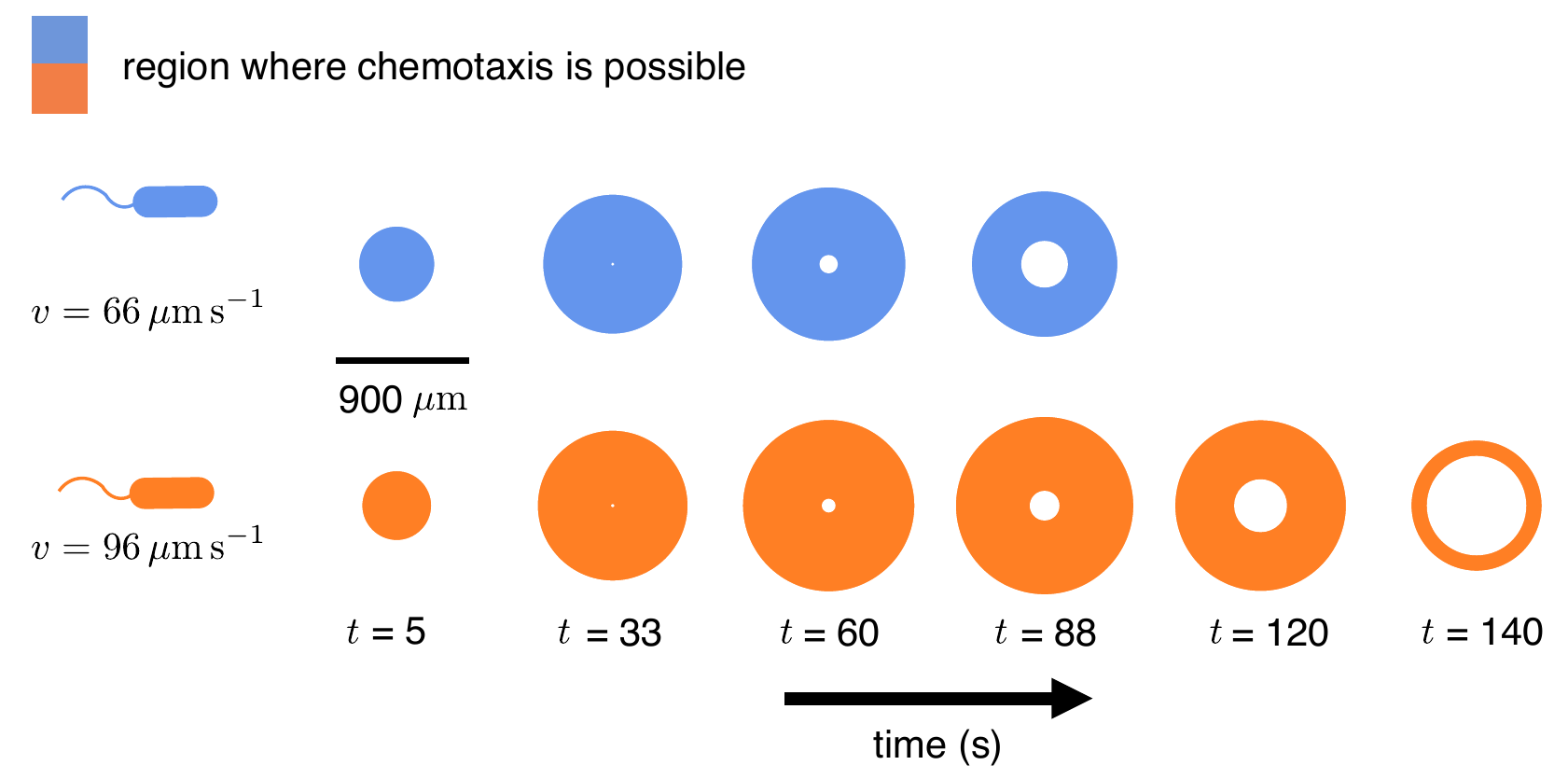}
\end{center}
\caption{Effect of swimming speed on the time evolution of the region where chemotaxis is possible. Colored regions show a two-dimensional cross-section of the region in which cells can resolve chemoattractant gradients (i.e. Conditions (\ref{eq:conditions}) are satisfied). Blue regions are those experienced by a cell travelling at a cruising speed typical of the bacterium \textit{V. coralliilyticus} ($\sim 66 \, \mathrm{\mu m \;s^{-1}}$). Orange regions are those experienced by a \emph{V. coralliilyticus} cell travelling at a high speed after initiating chemokinesis ($\sim 96\, \mu$m s$^{-1}$) \cite{garren:2013}. Other parameters as in Fig.~2. Note the \emph{blind spot} that forms at the centre of the region as the inner boundary of sensitivity, $r_i$, expands.}
\label{fig:kinesis}
\end{figure}

Effects of changes in speed may be substantial. For example, the coral pathogen \emph{Vibrio coralliilyticus} increases its speed by as much as 45\% when chemoattractant concentration is high \cite{garren:2013}. The temporal evolution of a chemoattractant pulse appears very different to a bacterium swimming at $66 \; \mathrm{\mu m \, s^{-1}}$ (typical cruising speed of \textit{V. coralliilyticus} and other \textit{Vibrio} spp.; Fig.~\ref{fig:kinesis}, blue regions) than it does to a bacterium travelling at speeds closer to $100 \; \mathrm{\mu m \, s^{-1}}$ (swimming speeds of chemokinetic \textit{V. coralliilyticus}\cite{barbara:2003,garren:2013}; Fig.~\ref{fig:kinesis}, orange regions).

\section*{Discussion}

Bacteria must cope with considerable noise and estimation bias when navigating dynamic chemical landscapes. The advantage conferred by an early response to chemical pulses suggests that there may be selection for high accuracy and sensitivity in the chemotaxis response \cite{taylor:2012,stocker:2012}. Our framework provides a means of studying how the basic components of bacterial navigation strategies (swimming speed, measurement time) and physical parameters (e.g., chemoattractant diffusivity, pulse size) influence when and where bacteria can perform chemotaxis. Expressions for the outer boundary of sensitivity, $r_o$ (Eq.~\ref{roapprox}), and the time after which gradients created by a pulse are no longer perceptible, $t^*$ (Eq.~\ref{tstar}), may prove particularly useful as they constrain the length and timescales over which bacteria can perceive individual chemical pulses. The relationship between the size of the pulse, pulse geometry, and the length and timescales over which the pulse is perceptible provides a basis for modeling more realistic environments where many pulses appear with characteristic sizes, geometries, and temporal statistics. For example an empirical estimate of typical inter-pulse-interval in, say, a marine environment \cite{stocker:2012}, can be compared to $t^*$ to determine whether the environment is highly granular or relatively homogeneous from the perspective of bacteria. For the canonical pulse geometries considered here (Eq.~(\ref{conc_radial})), the signal-to-noise ratio of the concentration ramp rate decays sharply far from the origin of a pulse (Fig.~2a, blue, red, and green curves). In particular, substituting  Eq.~(\ref{conc_radial}) into the expression for the SNR of $\hat{c_1}$ (r.h.s. of Eq.~(\ref{del})) shows that the SNR decays like a Gaussian for large $r$ ($ \mathrm{SNR} \propto \exp[-r^2/(8Dt)]$ for large $r$). This sharp transition in the SNR means that, near the outer boundary of sensitivity, there is a stark division between cells that have access to useful chemotactic information ($ r < r_o$) and cells that do not ($r > r_o$). Using $r_o$ to partition bacterial cells into subpopulations that are near and far from chemical pulses could greatly simplify models of bacterial competition and population dynamics in complex environments \cite{taylor:2012}. 

Our theory makes a number of predictions that could be tested with chemotaxis experiments. First, the theory predicts that for times $t < t^*$, the mean orientation of bacterial swimming trajectories outside the region $r_i < r <r_o$ should be unbiased. Because the conditions considered in this work correspond to an upper bound on sensory accuracy, the region within which cells exhibit biased motion may be a sub-region of $r_i < r <r_o$. A second prediction is that, for times greater than $t^*$, bacteria should not exhibit biased motion anywhere in the environment because each cell's estimate of the gradient will be dominated by noise, regardless of where it is located relative to the origin of the pulse. Again, because of the assumptions used to derive $t^*$, the observed time at which the average directional bias of a bacterial population drops to zero may be shorter than $t^*$. 

One of the implications of our model for temporal gradient sensing is that sensory acuity is intimately linked to swimming speed (Eq.~\ref{del}, Fig.~5). Because swimming at high speed is costly \cite{taylor:2012,berg:1977}, bacteria likely benefit by changing speed in an adaptive way, cruising at low speed in the absence of a chemical signal, and speeding up when concentration exceeds a threshold. The connection between speed and measurement accuracy may explain the counterintuitive observation that some species of marine bacteria swim at high speeds even near local maxima in chemoattractant concentration \cite{barbara:2003}; bias in the concentration slope estimate is high near local maxima (Fig.~\ref{noisy_grad}b). A cell cannot decrease bias by lengthening measurement time, but it can reduce bias by swimming faster, suggesting that bacteria may use chemokinesis to enhance {chemotactic accuracy} near the \emph{blind spot} that forms at the centre of spreading chemical pulses (Fig.~5 $t$ = 120 s, $t$ = 140 s; Supplementary Text). {More generally, our framework suggests that bacteria can improve chemotactic performance by using chemokinesis and chemotaxis in concert.} The hypothesis that bacteria initiate chemokinesis in response to absolute concentration to enhance sensitivity to gradients could be investigated by independently varying the concentration gradient and absolute concentration of a chemoattractant, for example using a microfluidic device \cite{son:2015}. 

Our framework uses fundamental limits on the accuracy of chemical sensing \cite{mora:2010,bialek:2005} to determine when and where chemotaxis is feasible, and provides a tool for modeling bacterial behaviour in more realistic dynamic environments. Importantly, it is agnostic to the details of bacterial movement patterns and chemosensory machinery and can therefore provide general principles that apply to the broad range of bacterial species in real ecological communities that navigate using temporal gradient sensing.

\section*{Acknowledgements}
This work was supported by Army Research Office Grants W911NG-11-1-0385 and W911NF-14-1-0431 to S.A.L., a James S. McDonnell Foundation Fellowship to A.M.H., a Swiss National Science Foundation postdoctoral fellowship to F.C., a Human Frontier Science Program Cross-Disciplinary fellowship to D.R.B., and a Gordon and Betty Moore Marine Microbial Initiative Investigator Award (GBMF3783) to R.S.

\bibliographystyle{vancouver}

\newpage
\section*{Supplementary Text}
\setcounter{figure}{0}
\renewcommand{\thefigure}{S\arabic{figure}}
\setcounter{equation}{0}
\renewcommand{\theequation}{S\arabic{equation}}
\newcommand{\textnew}[1]{{\color{blue}#1}}
\newcommand{\textcut}[1]{{\color{red}\sout{#1}}}
\newcommand{\textmod}[1]{{\color{orange}#1}}

\subsection*{Ramp rate and concentration slope estimation} 

\textbf{Estimating the ramp rate from a series of molecule absorptions.} Here we discuss constraints on the estimation of the ramp rate $c_1$ and the concentration slope $g$. Following \cite{berg:1977_supp,endres:2008_supp,mora:2010_supp}, we approximate a cell as an idealized measuring device: a sphere of radius $a$ that absorbs all molecules that come in contact with its surface. 

We begin by recalling relevant results of \cite{mora:2010_supp}. The average flux of molecules (molecules $\times$ time$^{-1}$) arriving at the surface of the sphere at position $\mathbf{x}$, time $t$ is $\langle I(\mathbf{x},t) \rangle = 4\pi D a C(\mathbf{x}, t)$ \cite{berg:1977_supp,mora:2010_supp} , which is equivalent in our notation to $\langle I(t) \rangle = 4\pi D a c(t) $. Again, the far-field chemoattractant concentration in the medium surrounding the sphere $c(t)$ can be linearized to $c(t) \approx c_0 + c_1 (t-t_0)$. The question discussed in \cite{mora:2010_supp} is: given a series of absorption times $\{ t_i\}, \; i=1, 2, ..., n$ measured during the interval $t_i \in (t_0 - T/2, t_0 + T/2)$, what is the minimum variance in the estimate of $c_1$ that a cell could possibly achieve? A natural tool for answering this question is the statistical framework known as Maximum Likelihood. Given a set of data (the time series $\{t_i\}$) and a generating model for those data -- in this case, that $c(t) = c_0 + c_1 (t-t_0)$, and absorptions are Poisson with rate $\langle I(t) \rangle = 4\pi D a c(t) $ -- one seeks values of the parameters of the generating model ($c_0$ and $c_1$) that maximize the probability, or ``likelihood", of the data. Maximum likelihood estimates are optimal in the sense that, as the number of observations becomes large, the variance of these estimators approaches a theoretical minimum variance for any unbiased estimator, which is given by the Cram\'er-Rao theorem \cite{rice:1995_supp}. This lower bound can be used to establish a bound on the accuracy with which cells can measure changes in concentration.

In the context considered in \cite{mora:2010_supp} and in our study, molecule absorptions are assumed to be independent Poisson events. Let  $t = 0$ be the time at which the pulse appears, and $t = t_0$ be a reference time marking the midpoint of the measurement interval $(t_0 - T/2, t_0 + T/2)$. The lengths of the time intervals between molecule arrivals, $\sigma_i = t_{i} - t_{i-1}$ \cite{asmussen:2007_supp} obey
\begin{equation}
\mathbb{P}(\sigma_i) 
= \langle I(t_i) \rangle \exp \left[ -\int_{t_{i-1}} ^{t_i} \langle I(s) \rangle ds\right],
\label{probsig}
\end{equation}
where $\sigma_1$ is defined as $t_1 - (t_0 - T/2)$ and the integral in Eq. (\ref{probsig}) is taken from $t_0 - T/2$ (the start of the measurement interval) to $t_1$ for $\sigma_1$. The probability of observing the set $\{t_i\}$ is

\begin{equation}
\mathbb{P}(\{t_i\}) =  \mathbb{P}(\{ \sigma_i\})  =   \prod_{i=1}^n \langle I(t_i) \rangle e^{-\int_{t_0 - T/2}^{t_0 + T/2} \langle I(t)\rangle dt}.  
 \label{lik2}
\end{equation}
By solving $\partial \log(\mathbb{P}[\{ t_i\})]/\partial \hat{c}_0 = 0$ and $\partial \log[ \mathbb{P}(\{ t_i\})]/\partial \hat{c}_0 = 0$, one can show that the values of $\hat{c}_0$ and $\hat{c}_1$ that maximize Eq. (\ref{lik2}) are:
\begin{equation}
\hat{c_0} = \frac{n}{4\pi D a T}
\label{c0hat}
\end{equation}
and
\begin{equation}
\hat{c_1} = \hat{c_0}\frac{\sum_i (t_i - t_0)}{\sum_i (t_i - t_0)^2},
\label{c1hat}
\end{equation}
where $n$ is the number of molecules absorbed during the observation interval. {Note that $\langle n \rangle \approx 4\pi D a c_0 T$ as the measurement interval becomes long, and as $n$ becomes large, 
$\sum_i (t_i-t_0) \approx \langle\sum_{t_i} (t_i - t_0) \rangle =  4 \pi D a\int_{t_0-T/2}^{t_0+T/2} [c_0 + c_1(t-t_0)] (t-t_0) dt = \pi D a c_1 T^3/3$ and 
$\sum_{t_i} (t_i - t_0)^2 \approx \langle \sum_{t_i} (t_i - t_0)^2 \rangle = 4 \pi D a \int_{t_0-T/2}^{t_0+T/2}[c_0 + c_1(t-t_0)] (t - t_0)^2 dt = \pi D a c_0 T^3/3$}, indicating that  maximum likelihood estimators, $\hat{c_0}$ and $\hat{c_1}$, are asymptotically unbiased, i.e., 
\begin{equation}
\hat{c_0} \rightarrow c_0 \qquad \mathrm{and} \qquad \hat{c_1} \rightarrow c_1 \qquad \mathrm{for \; large \;}n.
\label{c1hat_ass}
\end{equation}
{We derive Eq. (1) in the Main Text by calculating a lower bound on the variance of the ramp rate estimator, $\hat{c_1}$. The Cram\'er-Rao theorem states that the variance of $\hat{c_1}$ is bounded by the relation \cite{rice:1995_supp}:

\begin{equation}
\mathrm{var}(\hat{c_1}) \geq - \mathbb{E}\left[ \frac{\partial^2 \log(\mathbb{P}(\{t_i\};{c_1} ) )}{\partial {c_1}^2}\right]^{-1} = \mathbb{E}\left[ \sum_{t_i} \frac{(t_i - t_0)^2}{[ {c}_0 + {c}_1 (t_i - t_0)]^2} \right]^{-1}.
\label{crbound}
\end{equation}
Employing the assumption that $c_0 \gg c_1T$, and using $\sum_{t_i} [t_i - t_0]^2 \approx  \pi D a c_0 T^3/3$ as the number of absorptions becomes large implies 
\begin{equation}
\mathrm{var}(\hat{c_1})  \gtrsim \frac{c_0^2} {\sum_{t_i} [t_i - t_0]^2}  \approx \frac{3c_0}{\pi D a T^3},
\end{equation}
which is the relation given in Eq. (1) of the Main Text.

\textbf{Estimating the concentration slope in static and dynamic concentration fields.}} We are concerned with cells that use their estimate of the ramp rate $\hat{c_1}$ to estimate the spatial gradient in chemical concentration, which we refer to as the concentration slope. We also assume that the concentration field $C$ changes over time as a pulse spreads. In this setting, the cell must estimate the concentration slope $g$ along its path using some estimator $\hat{g}$ that can be computed from a series of observed absorption times. The concentration experienced by the cell can still be written $c(t) = c_0 + c_1 (t - t_0)$, but now $c_1 \approx gv + \partial C/\partial t$. If we begin by assuming $\partial C/\partial t = 0$, the maximum likelihood estimator for $g$ follows from the estimator for $c_1$:
\begin{equation}
\hat{g} = \hat{c_0}\frac{\sum_i (t_i - t_0)}{v\sum_i (t_i - t_0)^2}.
\label{ghat}
\end{equation}
In the limit of many molecule absorptions, $\sum_i (t_i - t_0) \approx \pi D a v gT^3/3$ and $\sum_i (t_i - t_0)^2 \approx \pi D a c_0 T^3/3$ (using the assumption that $c_0 \gg c_1T$),  which imply that $\hat{g}$ approaches the true concentration slope $g$ as the number of molecule absorptions becomes large (i.e., $\hat{g}$ is asymptotically unbiased). 

 When $\partial C/\partial t$ is not equal to zero and the cell is swimming at speed $v > 0$, the absorption time series $\{ t_i \}$ does not contain the information necessary to estimate both $g$ and $\partial C/\partial t$. This can be shown by combining Eq. (\ref{c1hat}) and (\ref{c1hat_ass}):

\begin{equation}
\hat{c_1} = \frac{n\sum_i (t_i - t_0)}{4\pi D aT \sum_i (t_i - t_0)^2} \rightarrow c_1 = gv + \partial C / \partial t.
\label{underdet}
\end{equation}
Equation~(\ref{underdet}) is clearly underdetermined; an infinite number of $g$ and $\partial C/\partial t$ value pairs can satisfy Eq. (\ref{underdet}). Without additional information, any estimator of the concentration slope $g$ will be biased. For instance, a cell could implement the maximum likelihood estimator $\hat{g}$ defined above to estimate the concentration slope in a dynamic environment. For $\partial C/\partial t \neq 0$, the sum $\sum_i (t_i - t_0) \approx \pi D a (gv + \partial C/\partial t) T^3/3$,  which implies that 
\begin{equation}
\hat{g} \rightarrow g + (\partial C/\partial t)/v.
\label{speed_bias}
\end{equation}
Equation~(\ref{speed_bias}) illustrates two important points: the bias in the concentration slope estimate (second term on the r.h.s. of Eq. (\ref{speed_bias})) is reduced as swimming speed increases; and this bias does not depend on measurement time $T$. One could propose alternative estimators for the concentration slope $g$ that satisfy Eq. (\ref{underdet}), but these basic conclusions are unchanged.

\subsection*{Dynamics of the outer boundary} 

\textbf{Derivation of the outer boundary, $r_o$.} To derive the time-scaling of the outer boundary $r_o$, we consider a cell that is travelling directly toward the origin of the pulse at speed $v$. {We define $r_o$ as the largest radius, $r$ that satisfies Eq. (4) in the Main Text. To approximate this value,  note that temporal changes in concentration are described by}
\begin{equation}
\frac{\partial C}{\partial t} = \left[ \frac{r^2}{4Dt^2}  -  \frac{N}{2t}\right]C,
\end{equation}
for the concentration profile studied in the Main Text, which implies that near $r = \sqrt{2 N Dt}$, temporal changes in the concentration field are small. {We assume $r_o$ is in this region and therefore neglect contributions of $\partial C/\partial t$ to the ramp rate measured by a swimming cell. This implies that the condition for the signal-to-noise ratio to rise above $\delta_0$ is $ -v \frac{\partial C}{\partial r} C^{-1/2} =  vr{C}^{1/2}/(2Dt) \geq \delta$. Solving for $r$ gives:

\begin{equation}
r = \sqrt{-4Dt \,{W}(-16Dt d^2)},
\label{outer_inner_rad_soln}
\end{equation}
where $d = \delta(4 \pi Dt)^{N/4} (4\sqrt{M}v)^{-1}$, and ${W}(\cdot)$ is the product log function. }In general, $M$ will be large so the argument of the product log function will be negative and close to zero (because $d$ is small). An approximation for the branch of the product log function that corresponds to $r_o$ in this regime is $W(x) \approx \ln(-x) - \ln(-\ln(-x))$ \cite{abramowitz:1964_supp}, which yields the approximation for $r_o$ given by Eq. (5) in the Main Text. 

\textbf{Derivation of the time when chemotaxis ceases, $t^*$.} The signal-to-noise ratio takes its maximum value at $r = \sqrt{4 D t}$. Near this radius the contribution of temporal changes in the chemical field to the cell's perceived ramp rate are small and, again, the signal-to-noise ratio is approximately proportional to $ -v \frac{\partial C}{\partial r} C^{-1/2}$. Solving for the time at which the maximum signal-to-noise ratio falls below threshold yields the expression for $t^*$ given by Eq. (6) in the Main Text.

\section*{Dynamics of the inner boundary.} 
The inner boundary $r_i$ is given implicitly by
\begin{equation}
\left| v g(r,t) + \frac{\partial C(r,t)}{\partial t}\right| C(r,t)^{-1/2} - \delta = 0,
\label{ri_equation}
\end{equation}
with $g = \partial C/\partial r$. Eq. (\ref{ri_equation}) has zero, one, or two positive roots. When this expression has no positive roots, cells travelling down the concentration gradient experience a signal-to-noise ratio of the ramp rate estimator that is below threshold $\delta_0$ everywhere. When this expression has one positive and one negative root, there is a maximum distance, beyond which cells travelling down the concentration gradient typically fail to detect a signal that is resolvable above noise, but any cell within this outer radius can typically resolve the ramp rate (Fig. 3 of the Main Text, early time). When Eq. (\ref{ri_equation}) has two positive roots, there exists an inner boundary, $r_i > 0$ within which, cells cannot resolve the ramp rate. This latter case is shown in Fig. 2a (dashed blue curve) and Fig. 3 inset in the Main Text.

\end{document}